# BrainKnow: Extracting, Linking, and Synthesizing Neuroscience Knowledge

Cunqing Huangfu*†, Kang Sun†, Yi Zeng*†‡, Dongsheng Wang*, Yuwei Wang*†, Zizhe Ruan*†

* Institute of Automation, Chinese Academy of Sciences, Beijing, 100190, China
† Center for Long-term Artificial Intelligence, Beijing, China
‡ Laboratory of Brain Cognition and Brain-inspired Intelligence Technology, Chinese Academy of Sciences, China

Corresponding author:
Cunqing Huangfu. Email: cunqing.huangfu@ia.ac.cn
Yi Zeng. Email:yi.zeng@ia.ac.cn

Cunqing Huangfu and Kang Sun contributed equally to the work.

# Abstract

The exponential growth of neuroscience literature presents a significant challenge for researchers seeking to efficiently access and utilize relevant information. To address this issue, we introduce the Brain Knowledge Engine (BrainKnow), an automated system designed to extract, link, and synthesize neuroscience knowledge from scientific publications. BrainKnow constructs a comprehensive knowledge graph encompassing 3,626,931 relationships across 37,011 neuroscience concepts, derived from 1,817,744 articles. This vast repository of knowledge is accessible through a user-friendly web interface, facilitating efficient navigation and data retrieval. BrainKnow employs advanced graph network algorithms, specifically Node2Vec, to enhance knowledge recommendation and visualization. This enables users to explore semantic relationships between concepts, predict potential new relationships, and gain a deeper understanding of the interconnectedness within neuroscience. Additionally, BrainKnow ensures real-time updates by synchronizing with PubMed, providing researchers with access to the most current information. BrainKnow serves as a valuable resource for neuroscience researchers, offering a powerful tool for exploring, synthesizing, and leveraging the vast and complex knowledge base of the field.

# Introduction

The pace of scientific publication in neuroscience is astonishing, presenting researchers with the formidable challenge of keeping up with developments in their fields. Structuring neuroscience knowledge by mapping the relationships between key research concepts within the domain can encapsulate specialized knowledge and highlight the interconnectivity inherent in the field. Neuroscience delves into the functions of the neural system by examining the interactions among cognitive functions, neurological disorders, and various levels of brain structure, including brain regions, neurons, genes, and neurotransmitters. These interactions form a cohesive knowledge network that enhances our understanding.Insights from multiple disciplines provide nuanced explanations of neuroscience phenomena. For example, studying depression involves analyzing the roles of specific brain regions, neurotransmitters, and genes linked to depressive disorders, the effectiveness of antidepressants, and the observed cognitive impairments in different depression subtypes. To comprehensively understand the mechanisms underlying depression, it is crucial to grasp the complex relationships among various neuroscience concepts. This includes identifying the target proteins, neurons, or brain regions affected by medications, evaluating the effects of specific genetic mutations on interregional connectivity, and clarifying how altered cognitive functions correlate with specific brain regions, neurons, neurotransmitters, or genes. These connections enhance our understanding of the root causes of specific phenomena and foster a deeper comprehension of the mechanisms at play. Building a knowledge network that maps the interrelations among neuroscience concepts can effectively summarize neuroscience knowledge and track the latest advancements in brain science research.

While there are existing contributions in this field, none completely realize this functionality. Academic search engines enable retrieval of papers based on specific keywords. However, the sheer volume of papers retrievable through keyword searches can total in the tens of thousands, and distilling this accumulated knowledge often requires considerable manual effort. Moreover, identifying precise search keywords can be challenging. Additionally, keyword-based searches typically only yield materials directly related to the search terms and do not support uncovering the relationships between these materials and information that is indirectly related.

Knowledge graphs and scientific databases serve as frameworks for organizing the intricate relationships among scientific concepts, facilitating researchers' access to information and enhancing their reasoning capabilities. In the era of big language models, knowledge graphs are increasingly crucial. A question-answering knowledge engine, powered by a large language model, can seamlessly integrate diverse knowledge aspects, offering users a convenient interface for interaction. However, large language models often encounter issues like hallucination when dealing with academic queries and perform suboptimally on less common, "long tail" topics. Employing Retrieval Augmented Generation (RAG) technology, as developed by (Lewis et al., 2020; Gao et al., 2024), knowledge graphs can underpin knowledge queries in generative large language models, effectively addressing issues related to hallucinations and coverage of rare topics. Additionally, knowledge graphs can support global statistical data for these models, providing comprehensive information on specific topics, such as detailing all neurons associated with a particular brain region.

Many knowledge bases extract relationship knowledge from scientific literature to construct knowledge graphs. KGBReF(Sun et al., 2021) extracted relations between 484 emotion-related concepts and 11,804 bacteria from 1,453 PubMed articles. Among these relations, they manually annotated 1,817 emotion-probiotic relations. KetPath(Liu et al., 2024) extracts relations from 2143 sentences from 2774 articles about the ketamine pathway. They trained a relation extraction model based on BioBERT to annotate the relation with one of the two labels: have relation, or do not have relation. PPKG(Liu et al., 2022) has constructed a knowledge graph focusing on pre-/probiotics and microbiota gut–brain axis diseases, employing a combination of manual curation and automatic knowledge extraction techniques.

Some knowledge bases related to brain science do not extract data directly from scientific literature but integrate data from other databases. LinkRBrain (Mesmoudi et al., 2015) integrates cognitive task data and gene expression information with specific brain structures. The dataset encompasses 451 cortical and 496 subcortical regions, 300 sensorimotor/cognitive tasks, and 21,000 gene expression profiles from multiple tasks. SPOKE(Morris et al., 2023) integrates relation knowledge from 41 databases. As far as we know, SPOKE is the only knowledge base related to brain science that is regularly and automatically updated.

Some scientific databases use knowledge discovery algorithms to further refine potentially new knowledge contained in the data. NeuroSynth(Yarkoni et al., 2011) employs text mining and meta-analysis techniques to autonomously establish precise associations between brain activity and a wide array of overarching cognitive states. BrainSCANr(Voytek and Voytek, 2012) analyzed the text of over 3.5 million scientific abstracts to find associations between neuroscientific concepts. These concepts encompass 124 brain structures, 291 cognitive functions, and 47 brain diseases. Utilizing the concept association indices, BrainSCANr generates relationship hypotheses automatically based on the "friend of a friend should be a friend" principle using the Jaccard similarity coefficient. MMiKG(Sun et al., 2023) extracted 1257 relations between Microbiota, Intermediate and Diseases manually. MMiKG, too, employs the Jaccard similarity coefficient to generate relationship hypotheses. Although the Jaccard similarity coefficient has successfully predicted the relationships between some concepts, it only considers the relationship information directly related to the two relevant concepts. The concepts in neuroscience form a complex network of relationships. Without considering the numerous relationships in the network, it is difficult to integrate effective information and make comprehensive and accurate predictions.

Bioteque (Fernández-Torras et al., 2022) extracts relations from various data sources and utilizes the Node2Vec algorithm to synthesize knowledge across different categories. Node2Vec involves "learning a mapping of nodes to a low-dimensional space of features that maximizes the likelihood of preserving network neighborhoods of nodes" (Grover and Leskovec, 2016a). In a network composed of nodes, the node vectors generated by Node2Vec can accurately represent the network relationships between nodes. That is, nodes that are close in location, have tight relationships, and share similar structural characteristics within the network will also have similar vectors. Node2Vec can effectively calculate the semantic relatedness between nodes in the networks, taking into consideration all relationships in the network.

The primary challenge in developing knowledge graphs and scientific databases is the labor-intensive process of data curation. Most of the knowledge engineering projects mentioned above suffer from infrequent updates following their initial release. Among these, SPOKE is the only database that receives regular updates. However, SPOKE does not directly extract knowledge from scientific literature, thereby missing many recent discoveries. The optimal approach for acquiring and maintaining a comprehensive repository of neuroscience knowledge is to extract information directly from academic papers, as these papers are consistently published and readily available online. Yet, the volume of information extracted from literature in the aforementioned databases is generally quite limited. This limitation often stems from the necessity for manual annotation of relationships or the restricted capabilities of automated extraction methods.

In knowledge graphs, integrating knowledge from multiple domains is a challenge that needs to be addressed. The knowledge within these graphs is highly complex. Whether it's about combining specific goals to acquire knowledge or predicting future knowledge based on current information, further integration of knowledge is required. Graph newwork algorithms like Node2Vec(Grover and Leskovec, 2016a) can summarize the overall relationship between concepts in a knowledge graph by producing embedding vectors. If two concepts have similar embedding vectors, it indicates that they have similar meanings within the graph. Bioteque employs the Node2Vec algorithm to create embeddings for concepts, which can be used to calculate their semantic relatedness, thus synthesis knowledge from various subdomains. However, the node embedding in Bioteque exclusively integrates concept relations within specific relation chains, such as [cell line]-[down/up-regulate]-[gene]-[down/up-regulate]-[cell line]. Consequently, the semantic relatedness between two concepts that do not belong to the same relational path might be incalculable. Furthermore, although Bioteque has made all embedding data accessible online, it lacks a user-friendly interface that facilitates convenient searching and exploration of brain knowledge.

We introduce the Brain Knowledge Engine (BrainKnow), a platform designed for the automated extraction of relationships among neuroscience concepts from scientific literature available on PubMed. The scientific literature dates back to the 1970s and is constantly updating. By February 2024, BrainKnow encompasses an extensive repository of 3,626,931 relationships and a range of 37,011 neuroscience concepts. The knowledge is extracted from 1,817,744 articles. These concepts span various domains, including brain diseases, cognitive functions, medications, brain regions, neurons, genes/proteins, pathways, and neurotransmitters. Access to this knowledge is facilitated through a user-friendly web interface, ensuring ease of retrieval and exploration. The knowledge repository is meticulously organized and offers multiple convenient querying methods. Importantly, BrainKnow ensures real-time updates by synchronizing with PubMed whenever new data becomes available.

# Results

## Relation Knowledge in BrainKnow

To systematically derive knowledge from textual sources, a comprehensive inventory of neuroscience-related concepts was assembled. This inventory encompasses a broad spectrum of neuroscientific elements, including brain diseases, cognitive functions, medicines, and fundamental brain components such as genes, neurons, neurotransmitters, pathways, and brain regions, as detailed in Figure 1A. The intricate process of curating these neuroscience concepts is elaborated in the "Systematic Compilation of Neuroscience Terminology" part of the Methods session.

The interconnections among these concepts encapsulate the majority of fundamental neuroscience knowledge. Cognitive function constitutes the cornerstone of neuroscience research. The majority of studies in this field are dedicated to elucidating the underlying mechanisms of cognitive processes. Brain diseases are often conceptualized as manifestations of cognitive function impairment. The exploration of cognitive functions typically entails an examination of the associations between these functions and the foundational elements of the brain, such as regions, proteins, and neurotransmitters. For instance, In the realm of learning and memory research, the inquiry extends beyond merely assessing the impact of genes, brain regions, and neurons. Researchers delve into the specifics of gene functionality within particular neurons, investigating how certain genetic mutations may lead to structural changes in brain regions, and encompasses an exploration of the distribution of specific neurons across various brain regions. This comprehensive approach allows for a more nuanced understanding of the intricate interplay between genetic factors and neuroanatomy in the context of learning and memory. Medicines are also pivotal, given their capacity to modulate neurotransmitter levels and activate neural pathways. The collected neuroscience concepts enable a comprehensive extraction of knowledge from scholarly articles. Our methodology involves analyzing the titles and abstracts of research papers from PubMed to distill this knowledge. We posit a relationship between two neuroscience concepts if they co-occur within the same sentence.

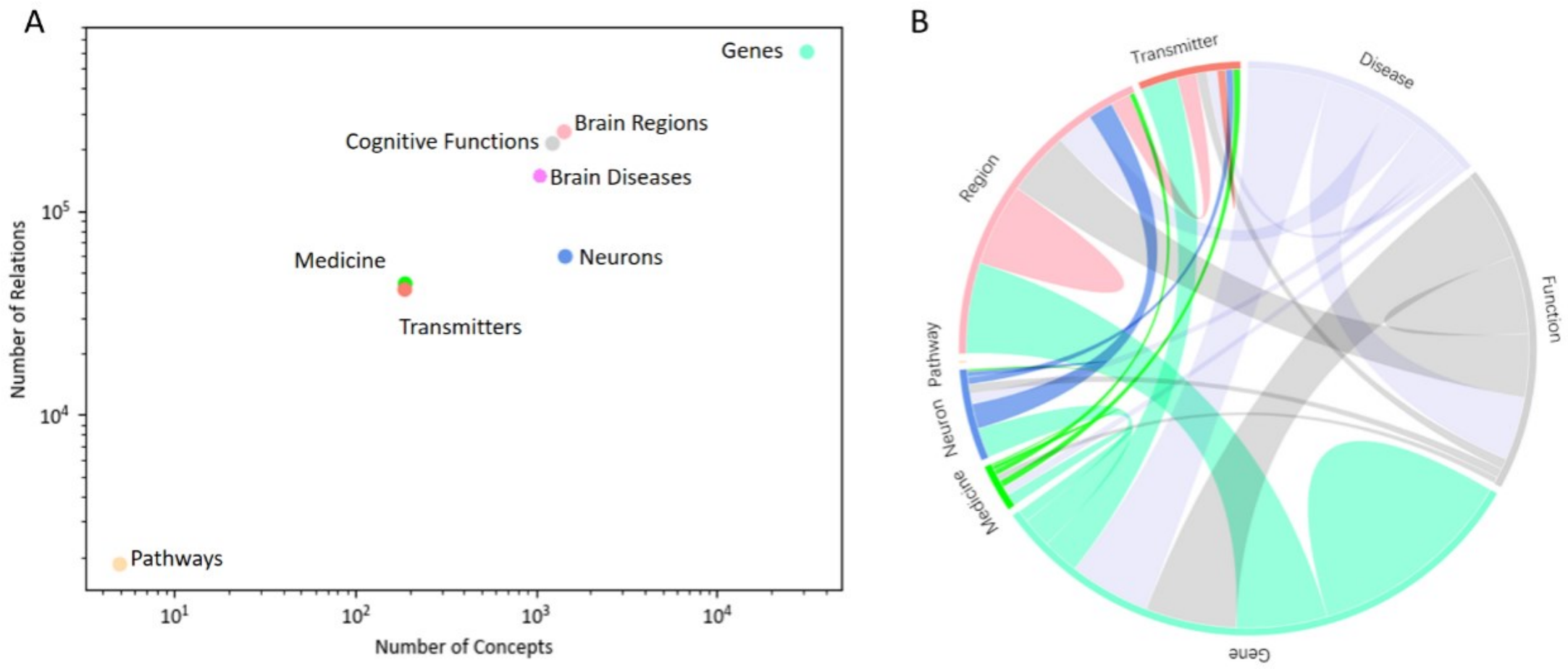

Figure 1 The Number of Relation Knowledge in BrainKnow

A: The number of each category of concepts and the number of relations of these concepts. B: The proportions of relationships between different types of concepts. The image was drawn using d3.js[1]. The width of the ribbon indicates the logarithm value of the number of the relations. The relations between genes, cognitive function, brain regions, and brain diseases contributed to most knowledge.

The proportion of relations between these concepts is demonstrated in Figure 1B. BrainKnow contains 3,626,931 relations between the concepts at the time that the article was written (February 2024). The statistics on the quantity of knowledge in the following article all come from this point in time. These relations are first organized according to the concepts. Figure 2A illustrates the page in BrainKnow that displays the relationships between two concepts, specifically, *set-shifting* and *the prefrontal cortex*[2]. We record the citation information of the article, the sentence, the extraction date, the publication date, and the speculated species about the article. The species of the article is speculated by searching species names in the title and abstract. All this knowledge is stored in the form of triples, with a total of 41,547,471 triples. As we can see, Figure 2A forms a comprehensive summarization of the relation between *set-shifting* and *the prefrontal cortex*. From old to new, researches gradually reveal the knowledge regarding the two concepts. Subscribing to this page enables researchers to track discoveries related to these two concepts. The species are inferred by scanning for species names within the title and abstract.

Figure 2B presents a comprehensive overview of the concepts related to *set-shifting* in the BrainKnow web interface. To explore the relationship between two specific concepts, users can interact with the 'Count' column. The parameters P(A|B) and P(B|A) are utilized to mirror the research prominence of each concept, illustrating the breadth of associations across various concepts. Specifically, P(A|B) quantifies the proportion of interactions between Concept A and Concept B relative to all interactions involving Concept B. For instance, *set-shifting* has 74 relations with *the prefrontal cortex* out of 631 total associations involving *set-shifting*, yielding a P(B|A) value of 0.117. Conversely, the P(A|B) value of *set-shifting* and *prefrontal cortex* is 0.0021, which indicates the lesser focus on *set-shifting* within the broader scope of *prefrontal cortex* research.

[1] https://observablehq.com/@d3/chord-diagram@416

[2] http://brain-knowledge-engine.org/relationType/function/set-shifting/region/prefrontal_cortex

Such analyses offer a foundational literature review regarding a specific concept. Additionally, BrainKnow can identify specific categories of concepts associated with a particular concept, as depicted in Figure 2C, which outlines the cognitive functions and brain regions linked to *Alzheimer's disease*.

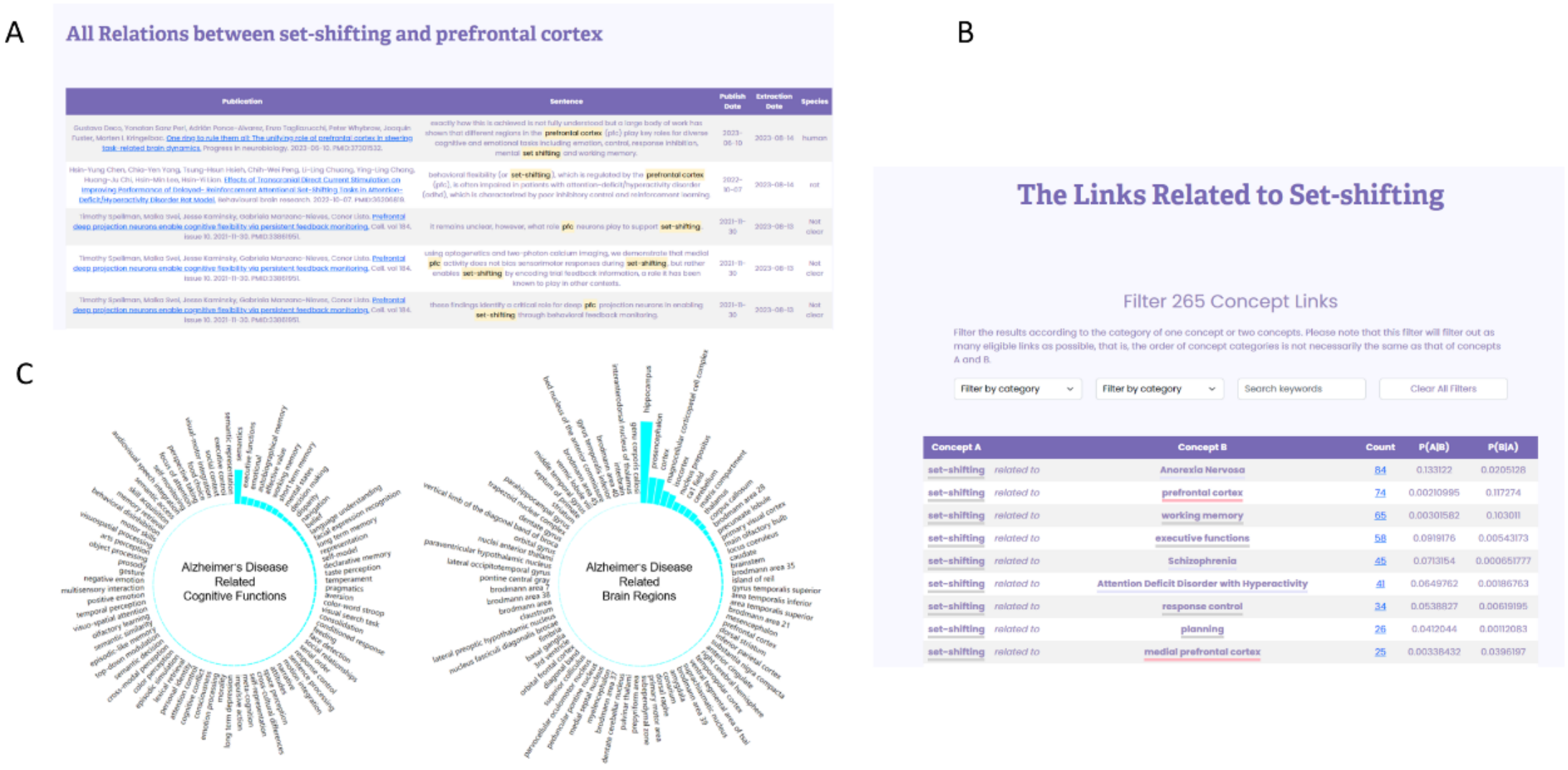

Figure 2 Illustrative Demonstration of Knowledge in BrainKnow
A: Relations between *set-shifting* and *prefrontal cortex* on the BrainKnow web interface. B: Knowledge related to *set-shifting* in the BrainKnow web interface. C: The cognitive functions and brain regions related to *Alzheimer's Disease*. The length of the ribbon indicates the logarithmic value of the number of relations.

## The Synthesize of Knowledge through Node Embedding

BrainKnow has amassed a substantial corpus of neuroscience knowledge and structured it through interrelations. However, the network of relationships among all the concepts is exceedingly intricate, rendering the precise retrieval of required knowledge quite challenging. Figure 3 delineates the relationships among the top 80 cognitive functions and brain regions, representing just a segment of the far more complex entire knowledge network. Each concept within this network is interconnected with numerous others. Certain concepts are closely linked to each other, thereby offering significant insights into the mechanisms underlying specific phenomena. Some concepts, while not directly related, share multiple common neighbors, highlighting their relevance in comprehending the overarching concept. Consequently, there is a pressing need for a methodology that can comprehensively synthesize the relation knowledge between concepts to accurately extract pivotal knowledge pertinent to one or several concepts.

BrainKnow employs the Node2Vec algorithm (Grover and Leskovec, 2016a) to integrate the overall association information between nodes in the knowledge network. This algorithm operates under the premise that an observer traverses the network from one node to another, with the travel probability being positively correlated to the weight of the relationship between nodes. As the traversal progresses, the trajectory of the observer is recorded. Prolonged travel and multiple iterations over all nodes ensure that this trajectory accurately represents the relative positions of the nodes within the network. The frequent appearance of two nodes within this trajectory implies a close relationship between them. This trajectory data is then input into a neural network designed to learn the co-occurrence of nodes along the trajectory. The penultimate layer of the neural network output for each node serves as its representation. Nodes that exhibit similar representations are deemed closely related within the network. To assess the efficacy of node embeddings in representing the relationships between nodes, the Area Under the Receiver Operator Characteristic Curve (AUROC) value is calculated to determine the correlation between embedding similarity and node relationships. An achieved AUROC value of 0.93 suggests a high proficiency of node similarity in accurately predicting node relationships(Figure 4A).

Figure 4B displays the top 20 concepts that are most semantically related to *set-shifting*. The specific computational methods are described in 'Querying Semantically Related Concepts' of the 'Methods' section. It is evident that concepts such as *prefrontal cortex*, *working memory*, *executive control*, *executive functions*, and *response control* share close semantic relationships. These concepts are particularly pertinent for users interested in exploring the domain of *set-shifting*. Notably, *schizophrenia* is associated with *set-shifting* through 45 relationships, as depicted in Figure 2B. However, given that *schizophrenia* does not demonstrate close semantic ties to other concepts related to *set-shifting*, its semantic relevance to *set-shifting* is relatively limited. This analytical mechanism aids users in filtering out broadly related concepts, such as *Alzheimer's disease* or *schizophrenia*, thereby enabling the retrieval of concepts that provide an accurate and succinct summary of knowledge directly pertinent to the queried concept.

BrainKnow also possesses the capability to retrieve concepts that are semantically related to multiple entities. This feature enables users to refine the scope of their information retrieval. Researchers seeking to delve deeper into this correlation would require a more comprehensive set of information. Figure 4C presents the 20 concepts most semantically related to both *set-shifting* and *the orbital frontal cortex*. The inactivation of *the orbital frontal cortex* has been demonstrated to affect *set-shifting* (Ghods-Sharifi et al., 2008). In contrast to Figure 4B, Figure 4C includes concepts elucidating the relationship between *set-shifting* and *the orbital frontal cortex*. For instance, the concept of *reward* significantly impacts *executive functions* like *set-shifting* (Pearce et al., 2018; Lertladaluck et al., 2020; Chan et al., 2022) and is intricately linked to *the orbital frontal cortex* (Rolls et al., 1999; Rolls, 2023; Dabrowski et al., 2024). In cases where direct evidence of relationships between *set-shifting* and *orbital frontal cortex* is not sufficient, the semantic relevance retrieved through the semantic relatedness of these two concepts provides us with valuable reference material. This enables us to gain a more convenient understanding of the complex relationship between the two concepts. This approach of adding more concepts to the search facilitates the retrieval of more precise results and progressively zeroes in on the core information about the specific inquiry.

On this basis, users can further explore this research topic within BrainKnow. *Set-shifting* is the ability to flexibly adjust behavioral strategies to adapt to new environments or task demands. This complex, high-level cognitive function relies on many other cognitive abilities as its foundation. Various brain regions associated with *set-shifting* may support its function through different cognitive processes. BrainKnow helps us quickly understand the diverse aspects of the relationship between different brain regions and *set-shifting*. Here, we compared the distinct relationships between the *orbital frontal cortex* and the *prefrontal cortex* with *set-shifting*. We conducted searches in BrainKnow using the keywords "*set-shifting* + *orbital frontal cortex*" and "*set-shifting* + *prefrontal cortex*" to find the most relevant cognitive functions(Figure 4D,4E). By comparing the results, we discovered that the *orbital frontal cortex* is more associated with reward-related functions (*reward expectation*, *reward learning*, *pleasantness*, *reward*), whereas the *prefrontal cortex* is more associated with memory-related functions (*working memory*, *spatial working memory*, *memory updating*). This aligns with our general understanding that the *prefrontal cortex* is more involved in short-term memory, decision-making, and attention regulation, while the *orbital frontal cortex* is more related to reward, emotion, and motivation. In figure 4F, we can clearly see the network of relationships among the aforementioned concepts. All the relationships between these concepts can be viewed in BrainKnow, providing solid evidence to support the highly abstracted data visualization results.

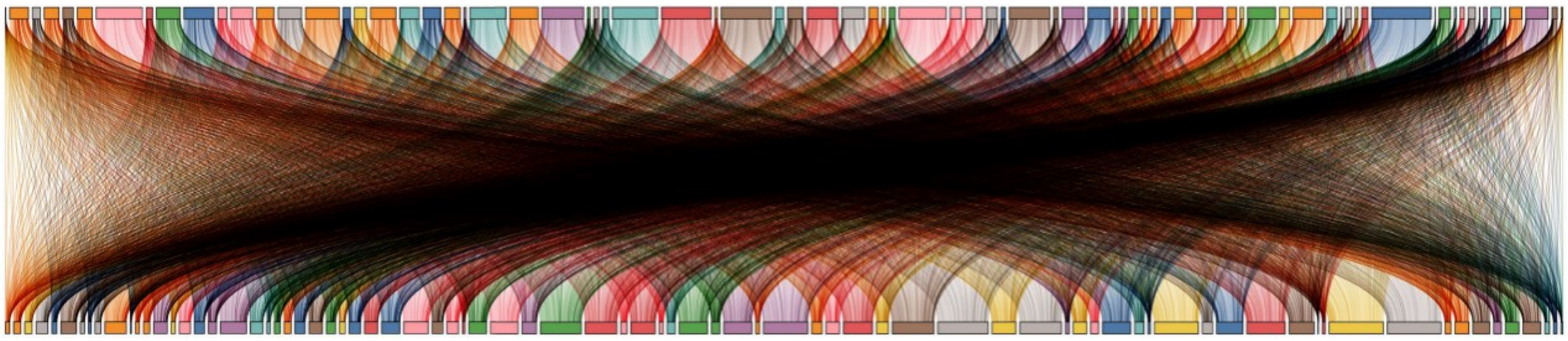

Figure 3 The Relations between the Top 80 Cognitive Functions and Brain Regions.
The widths of the ribbons indicate the logarithmic value of the relation count.

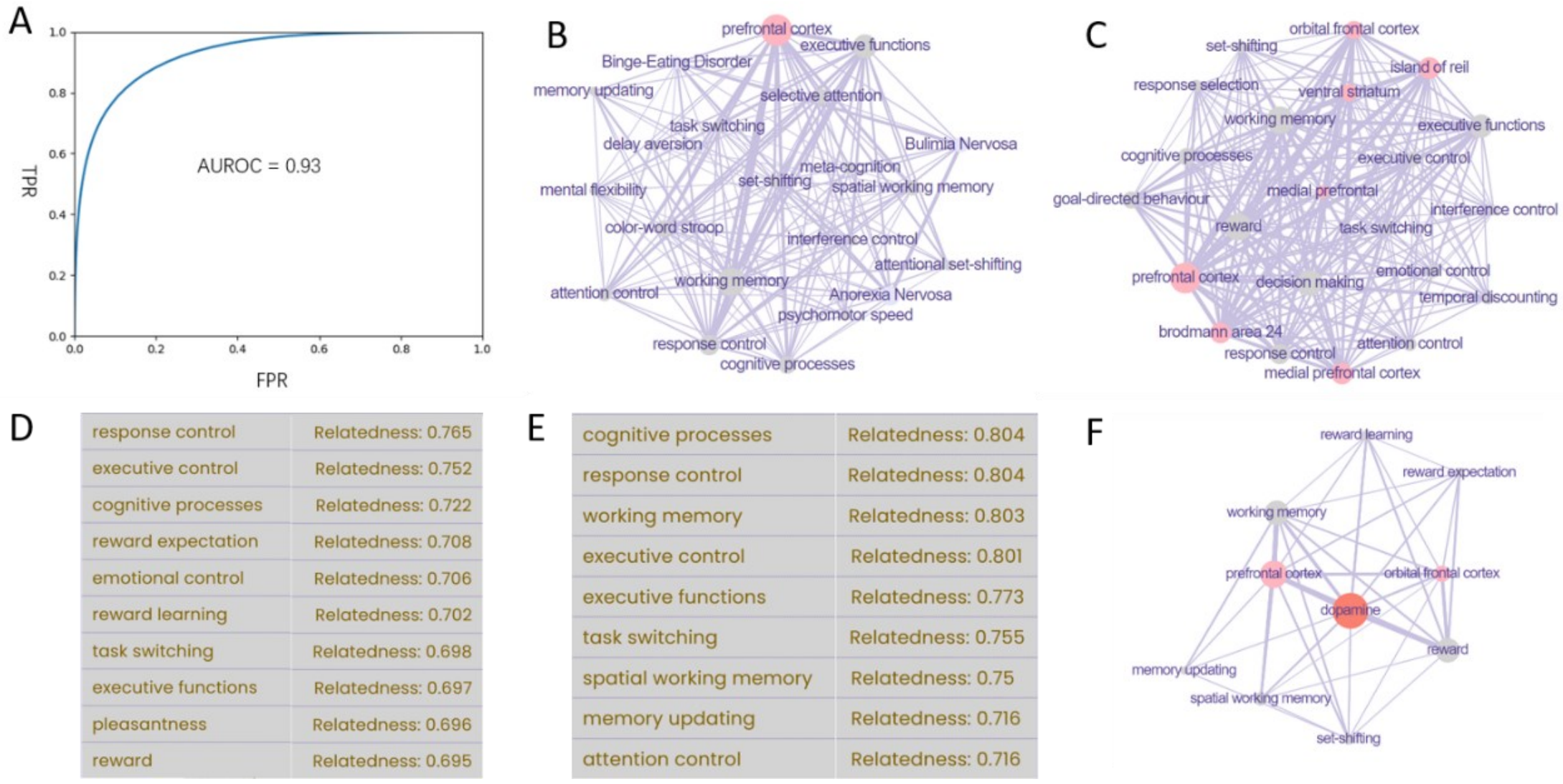

| | |
|---|---|
| response control | Relatedness: 0.765 |
| executive control | Relatedness: 0.752 |
| cognitive processes | Relatedness: 0.722 |
| reward expectation | Relatedness: 0.708 |
| emotional control | Relatedness: 0.706 |
| reward learning | Relatedness: 0.702 |
| task switching | Relatedness: 0.698 |
| executive functions | Relatedness: 0.697 |
| pleasantness | Relatedness: 0.696 |
| reward | Relatedness: 0.695 |
| cognitive processes | Relatedness: 0.804 |
| response control | Relatedness: 0.804 |
| working memory | Relatedness: 0.803 |
| executive control | Relatedness: 0.801 |
| executive functions | Relatedness: 0.773 |
| task switching | Relatedness: 0.755 |
| spatial working memory | Relatedness: 0.75 |
| memory updating | Relatedness: 0.716 |
| attention control | Relatedness: 0.716 |

Figure 4 Node Embeddings Can Accurately Integrate Relevant Knowledge
A: The AUROC curve validates the correlation between node embedding similarity and node relation. B: The 20 concepts that are most semantically related to *set-shifting* in BrainKnow. C: The 20 concepts that are most semantically related to *set-shifting* and *orbital frontal cortex* in BrainKnow. D: Cognitive functions that are most semantically related to *orbital frontal cortex* and *set-shifting*. E: Cognitive functions that are most semantically related to *prefrontal cortex* and *set-shifting*. F: Relations between *reward*, *memory*, *prefrontal cortex*, *set-shifting,* and *orbital frontal cortex*.

To showcase the comprehensive knowledge within BrainKnow, we utilized the TSNE algorithm to reduce the vector representations of all concepts to two dimensions, as illustrated in Figure 5. In the graph, we can observe that concepts have formed distinct clusters, each representing specific research domains. Different levels of brain structures are generally distributed continuously. Areas densely populated with brain regions are connected to areas containing neurons, which in turn link to areas where neural transmitters are located. The distribution of genes is quite extensive, demonstrating that genes are fundamental to the essential mechanisms of life activities. Brain regions are primarily concentrated in two areas. One area, located in the lower left corner, clusters together with cognitive functions and some brain diseases. This research area primarily investigates the relationships between brain region activity, cognitive functions, and brain diseases, often using techniques such as fMRI. Another concentration of brain regions targets the interrelations among brain regions, genes, and neurons as its research focus. Neural transmitters and some medicines are grouped together, consistent with many medicines acting as antagonists or activators of neural transmitter receptors. Brain diseases are primarily divided into two clusters: one in the lower left corner, closely associated with cognitive functions and brain regions; the other is positioned more centrally to the right, where the main impacts of these brain diseases are related to infectious brain diseases and traumas rather than specific cognitive functions.

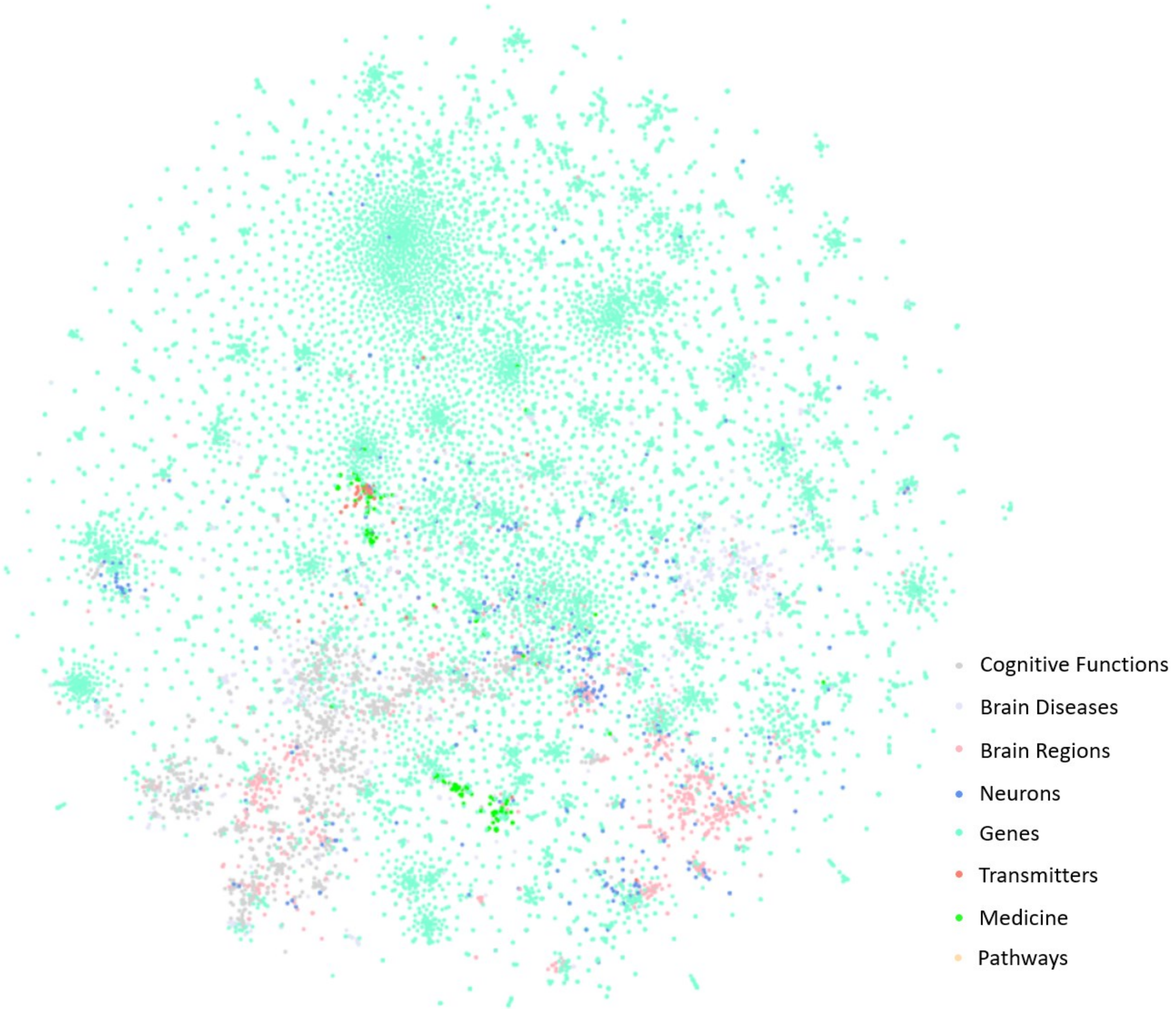

Figure 5 Distribution of Knowledge Concepts in BrainKnow Visualized Using t-SNE

## Predicting the relationship between concepts

The semantic relatedness between concepts can be employed as a predictive tool for potential new relationships. This is because node embedding encapsulates the overall connectivity between nodes. If two nodes in the network display similar embeddings yet lack direct connections, it suggests the existence of numerous indirect relationships between them. Consequently, within BrainKnow, if two concepts exhibit highly similar embeddings while do not have direct relationships, it strongly implies a likelihood of their relatedness. For instance, *shank3*, a gene known for encoding a postsynaptic protein in the *dopamine synapse*, influences the *dopamine pathway* and, consequently, various cognitive functions related to it. *Shank3* is notably recognized for its association with *autism*. Figure 6A delineates the cognitive functions that are semantically related to, but not directly connected with, *shank3*. It is observed that most of these functions pertain to social activities. Given that *shank3* mutations are involved in the *dopamine pathway*, which plays a role in social activities, and its strong correlation with *autism*, it is plausible to infer that *shank3* is related to a multitude of socially related cognitive functions, despite the lack of direct evidence. Figure 6B displays the top 20 concepts that are most semantically related to both *shank3* and *the theory of mind*. These concepts, encompassing genes, cognitive functions, and diseases, form an intricate network that facilitates the tracing of detailed knowledge relationships.

To evaluate the overall accuracy of predicting new relationships, we attempt to forecast recently extracted relationships using historical data. Specifically, we identify all relationships that emerged post-2020 and assess whether they could be predicted by node embeddings trained on data available from 1970 to 2020. For each relationship between two concepts (concept A and concept B), we calculate the most semantically related yet not directly connected concepts for both concept A and concept B. We then document the rank of a concept in the list of most related concepts for the other concept. If neither concept appears among these related concepts, the relationship is deemed unpredictable. Conversely, if both concepts are present in each other's list of most related concepts, we record the lower of the two ranks. Relationships involving at least one concept not present in the pre-2020 data are excluded from consideration.

Figure 6C illustrates the count of successfully predicted relationships, categorized by their rank positions ranging from 1 to 40. An increased level of semantic relatedness between two concepts correlates with a higher likelihood of unveiling new relationships between them. Notably, 164 relationships are predicted using the most semantically related yet not directly connected concepts. Among the 24,967 newly emerged relations, 2,250 are predicted from rank 1 to 40. This proportion may seem small. However, it is pertinent to acknowledge that the predictability of relationships is contingent upon certain specific conditions. High predictability is generally feasible when two concepts are part of a closely interconnected network of relationships. It is observed that many discoveries do not conform to this stipulation.

A

| Predicted Cognitive Function | Semantic Relatedness |
|---|---|
| joint attention | 0.834 |
| social attention | 0.819 |
| direct gaze | 0.803 |
| socialization | 0.787 |
| social competence | 0.778 |
| mind perception | 0.771 |
| theory of mind | 0.764 |
| imitation | 0.759 |

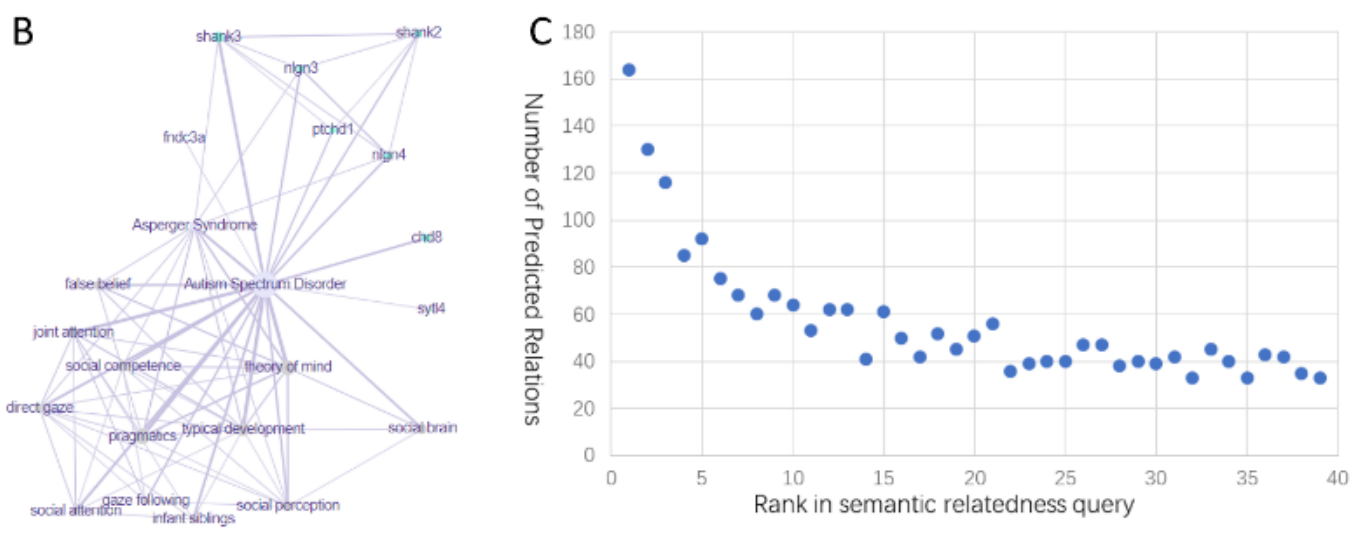

Figure 6 Node Embeddings Can Effectively Predict New Relation
A: The most semantically related, but not directly related cognitive functions of gene shank3 in BrainKnow. B: The 20 most semantically related concepts of gene shank3 and theory of mind. C: Node pairs with similar representations are more likely to form new relations. The x-axis denotes the rank of a predicted concept within the list of most semantically related concepts for another concept. The y-axis represents the count of predicted relationships at each specific rank of semantic relatedness. The quantity of predicted relationships at each rank position is proportional to the likelihood of successful prediction at that rank, considering that the total number of predicted relationships is consistent across each rank position.

# The Design of the Website Interface

The user interaction in scientific databases is a complex issue. As a knowledge engine, BrainKnow integrates knowledge based on relationships between concepts. During the interaction, users should be able to find the knowledge they need by using various relationships between concepts as clues. Figure 7A illustrates the various pathways through which users acquire knowledge when using BrainKnow. Firstly, users can select concepts using text search (Figure 7B) or from the concept category (Figure 7C). Clicking on concepts allows users to select them and add them to the list of selected concepts (Figure 7D). The selected concepts can be used to search for semantically similar concepts in the Relation Graph (Figure 7E) or directly retrieve related knowledge (buttons below Figure 7D). Concepts obtained through each of these methods can be further added to the list of selected concepts. This approach enables users to utilize the associations between every type of knowledge within BrainKnow for their queries.

BrainKnow, like PubMed, is updated daily. Website updates occur at 00:30 Beijing time. The entire update process, which includes downloading, knowledge extraction, data statistics, and node vector training, typically requires approximately 3 hours to complete. Throughout this period, the old version of BrainKnow continues to operate, ensuring the continuity of website services.

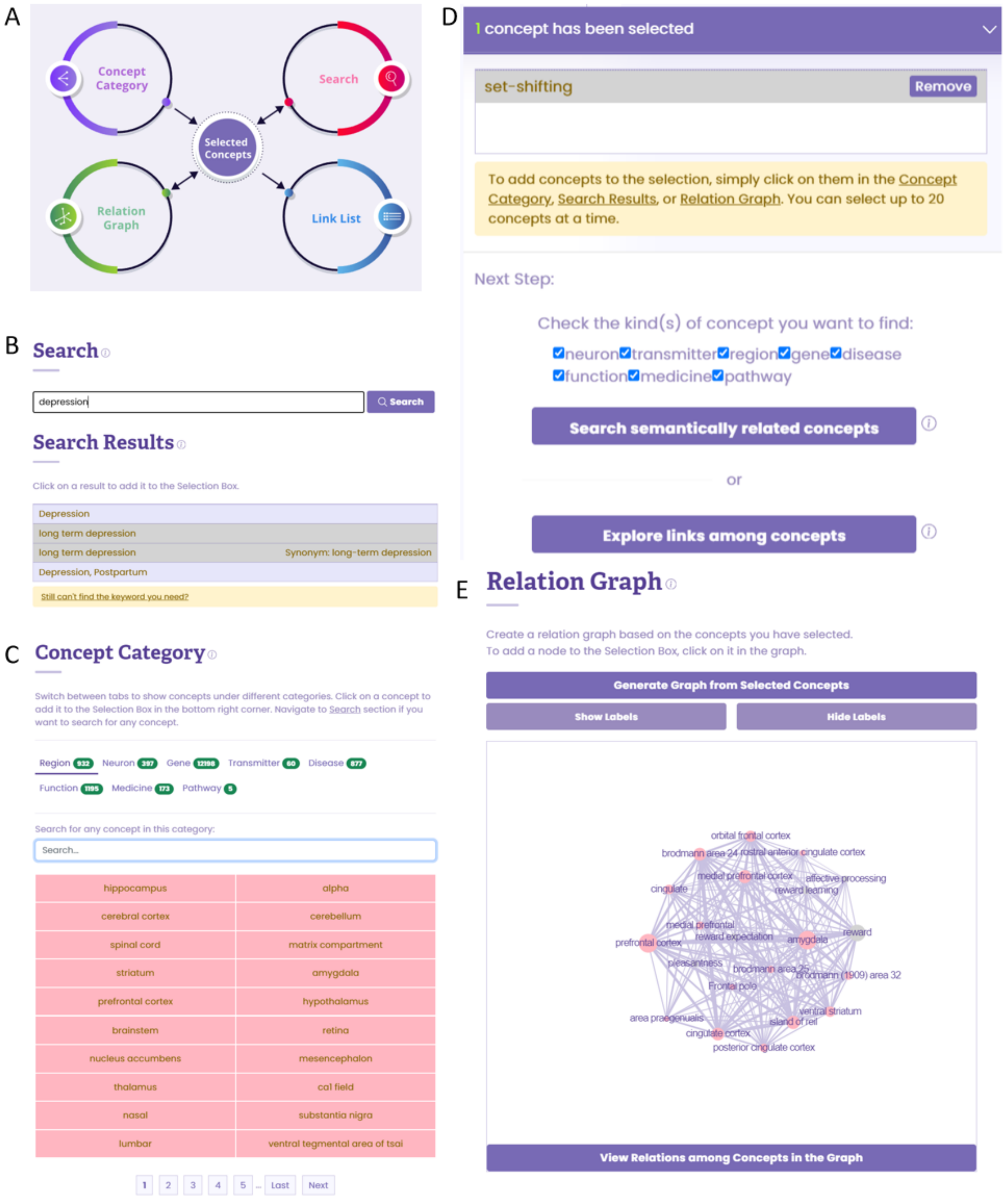

Figure 7 The Web Interface

A: The various pathways through which users acquire knowledge. B: Text search. C: The concept category. D: Selected concepts. E: Relation graph.

# Conclusion

BrainKnow provides researchers with a real-time updated and well-synthesized knowledge engine for neuroscience. Researchers can not only access relevant knowledge in their specific fields through various methods but also predict knowledge through the semantic relevance between concepts. Knowledge extraction from literature and statistical integration of the extracted knowledge provide researchers with a complete knowledge landscape.

# Discussion

The primary significance of BrainKnow lies in its use of vector space to offer users limitless possibilities for customized information filtering. The relationships between different pieces of knowledge are extremely complex. Researchers may need to explore the principles of the same issue from various perspectives, each with a different focus. Relying solely on keywords, topics, or research fields to obtain literature cannot meet these needs. BrainKnow creatively maps knowledge into a vector space and employs multiple methods to retrieve and present information based on their relative positions in this space.

Generative large language models have demonstrated significant potential in the specialized field of knowledge engineering. A series of advancements in research on large language models (Brown et al., 2020; Touvron et al., 2023; Vaswani et al., 2023; Radford et al., n.d.; Raffel et al., n.d.) has enabled more sophisticated interactions with intelligent agents using natural language, facilitating the acquisition of knowledge from diverse sources. Previously, the integration of knowledge across multiple domains was a complex challenge, with different pieces of knowledge siloed in separate databases, each requiring unique retrieval methods.

However, employing large language models directly as knowledge engines can present challenges such as hallucinations and the long-tail effect. Hallucinations in this context refer to the models generating plausible yet unfounded responses. The long-tail effect describes the models' decreased performance when addressing more obscure queries. To mitigate these issues, one strategy involves combining the natural language understanding capabilities of large language models with specific textual materials or knowledge graphs, thus grounding the models' responses in verifiable data. For instance, the Retrieval Augmented Generation (RAG) technique (Lewis et al., 2020; Gao et al., 2024) utilizes text and related materials to help large language models deliver more accurate answers. BrainKnow, with its detailed mapping of relationships between brain science concepts, can provide empirical grounding to help prevent hallucinations when integrated with a large language model. Furthermore, BrainKnow's extensive database of statistical relationship information spans a broad spectrum of niche research areas, effectively addressing the long-tail problem encountered by large language models in responding to specialized queries.

In BrainKnow, we did not differentiate between categories of relationships during information extraction for several reasons. Firstly, the accuracy of relationship extraction algorithms has been limited, and to improve accuracy, specific natural language processing models need to be trained or fine-tuned for each type of relationship. The number of relationship types among all neuroscience concepts is too numerous, making it impractical to train a model for each type. General-purpose relationship extraction algorithms like OpenIE(Dutta et al., 2014, 2015; Martinez-Rodriguez et al., 2018) would have even lower accuracy. Many knowledge engineering projects are greatly limited in scale due to the need to handle the details of relationships. Moreover, without the original context, it is challenging to describe the relationships between some concepts in a standardized manner. The degree of relationship determination often varies significantly in different specific contexts. Storing this information in a knowledge graph in a standardized form for reasoning would lead to many errors. Furthermore, with the rise of large language models, capturing semantic details has become easier. The detailed semantic relationships between concepts can be handled by generative large language models without the need to first organize them into a knowledge graph.

Looking ahead, BrainKnow is poised to advance in two primary directions. Firstly, we plan to integrate it with generative large language models to enable users to interact with the system using natural language. This integration will leverage Retrieval Augmented Generation (RAG) technology to incorporate information in a timely and organic manner. Secondly, BrainKnow does not differentiate between the types of relationships between concepts, which limits the amount of information contained in the knowledge graph. Large language models hold significant potential to enhance the performance of knowledge extraction algorithms. These models can not only improve the accuracy of knowledge extraction but also simplify the development process required for extracting knowledge in specific domains(Wadhwa et al., 2023). With the aid of these technologies, BrainKnow aims to construct more detailed and accurate knowledge graphs.

# Methods

## Systematic Compilation of Neuroscience Terminology

The list of brain diseases has been derived from the Medical Subject Heading (MeSH) taxonomy (Lipscomb, 2000), specifically, from categories C10 (Nervous System Diseases) and F03 (Mental Disorders). The compilation of cognitive functions was extracted from research keywords found in PubMed Central research articles (https://www.ncbi.nlm.nih.gov/labs/pmc/), with the methodology detailed in (Huangfu et al., 2020). A Word2Vec model was trained on a comprehensive corpus of full-text articles sourced from PubMed Central. Following training, research keywords were clustered using the Agglomerative Clustering algorithm based on their node vectors. Clusters containing cognitive functions were selected, and synonyms were manually organized, while invalid terms were excluded.

The list of medicines consists exclusively of neuronal receptor agonists and antagonists, and its composition is based on (Romano, 2019). Gene and protein terminology has been extracted from the Gene Ontology (GO) resource (Ashburner et al., 2000; The Gene Ontology Consortium et al., 2021), specifically from ontologies within the following categories: GO:0007212 (Dopamine Receptor Signaling Pathway), GO:0023041 (Neuronal Signal Transduction), GO:0098926 (Postsynaptic Signal Transduction), GO:0098928 (Presynaptic Signal Transduction), GO:0007210 (Serotonin Receptor Signaling Pathway), and GO:1990089 (Response to Nerve Growth Factor).

Neuron terminologies have been curated from the NeuroMorph database

(https://neuromorph.epfl.ch/) (Jorstad et al., 2015, 2018). Brain region nomenclature is drawn from both the NeuroMorph database and the Brain Atlas project (http://atlas.brainnetome.org/download.html), as well as Supplementary Table 1 (Oh et al., 2014). The neurotransmitters list is based on the compilation provided by (Cherry and Lakhan, 2021). Additional synonyms for neurons and brain regions have been sourced from NeuroLex (Larson and Martone, 2013) and incorporated into the lexicon. It is noteworthy that several concepts have been included based on private communication in addition to the aforementioned sources.

All terminologies can be acquired through the ***Search*** section on the index page of BrainKnow. All terminologies that have at least one relational information in BrainKnow can be acquired through the ***Concept Category*** section on the index page. All terminologies can also be acquired in the supplementary data.

Before utilizing these concepts for knowledge extraction purposes, a meticulous manual verification process is undertaken to ensure the utility, precision, and comprehensiveness of the extracted relationships. Concepts that carry the potential for introducing errors or inaccuracies are systematically excluded from the concept list. To illustrate, consider the abbreviation 'dId,' which corresponds to 'dorsal dysgranular insula.' However, 'dId' also coincides with the auxiliary verb 'did.' The inclusion of this abbreviation in the concept list would result in the extraction of a substantial volume of erroneous relationships. Given the imperative of maintaining the precision of relationship extraction, the incorporation of numerous concepts into the list without a rigorous manual validation process is precluded, thereby constraining the scale of the knowledge graph. In forthcoming iterations, a novel strategy will be adopted to diversify the sources of concepts.

## Knowledge Extraction

Concept relationships are extracted from articles retrieved from PubMed. Files containing newly published articles are dayly obtained from the following URL: https://ftp.ncbi.nlm.nih.gov/pubmed/updatefiles/. If new material has been updated, the system will undergo the upgrade process. While neuroscience articles are not explicitly distinguished from other articles within the dataset, the majority of concepts present in the lists are inherently related to neuroscience. Consequently, the knowledge extracted from these articles primarily pertains to the field of neuroscience. By the time the article is written (February 2024), BrainKnow contains knowledge extracted from 1,817,744 articles.

The extraction process starts with the retrieval of title and abstract texts from the data files, which are then preserved for subsequent information extraction activities. A relationship is recorded whenever two concepts are identified within the same sentence. Notably, the recognition process is case-insensitive. In cases where a concept consists of multiple words, such as 'Parkinson's disease,' the concept is tokenized into individual words utilizing the 'word_tokenize()' function. It should be acknowledged that some concepts may yield no relation extraction results.

When extracting relationships, it becomes time-consuming to search for a multitude of keywords within a large corpus of text. Conventionally, in the process of locating a specific keyword within a sentence, a straightforward linear scan from the beginning to the end suffices. However, when confronted with the task of matching tens of thousands of keywords against an extensive corpus, the corpus must be subjected to repeated scans, resulting in significantly reduced efficiency. To address this challenge, we employ a segmentation approach by breaking down sentences into individual words and subsequently verifying their presence against a keyword hashtable. Textual content is segmented into individual sentences using the 'sent_tokenize()' function provided by the Natural Language Toolkit (NLTK) package. Subsequently, sentences are further segmented into individual words employing the 'word_tokenize()' function from NLTK. This methodology ensures that the corpus is scanned only once, mitigating the computational burden.

For phrases, we employ a similar technique by disassembling the phrases into constituent words and constructing a dictionary tree for efficient search operations. The NLTK's 'sent_tokenize' function is utilized to segment both the corpus and phrases into individual words, facilitating the implementation of this strategy.

## Node Vector Training

The protocol for node vector training is outlined as follows: The entire knowledge base is transformed into an undirected graph, where each node represents a concept, each edge signifies a relationship, and the quantity of relationships between a specific pair of concepts is denoted as the weight associated with the edge. The model parameters used align with those examined in the original research paper on Node2Vec by (Grover and Leskovec, 2016b). The training parameters are introduced as follows: The dimension of the representation vector is 128, the walk length is 80, the number of walks is 18, the training epoch number is 10, the window size is 16, and the p and q parameters which control the walking behavior are both 0.25. The training process is executed through the utilization of the Python package Node2Vec[3].

## Querying Semantically Related Concepts

The process for querying semantically related concepts using node vectors is described as follows: During the query process, the vectors corresponding to the selected concepts are summed and then normalized to obtain a vector with a unit norm. Next, the search for similar concepts is performed using the model.wv.similar_by_vector_consmul() function, which is a key component of the Gensim Word2Vec model framework, developed by(Radim Řehůřek and Petr Sojka, 2010). The result of this query procedure is the retrieval of the top 20 concepts exhibiting the highest level of semantic relatedness.

[3] https://github.com/eliorc/node2vec

# Data Availability

All features and content are accessible via the URL http://brain-knowledge-engine.org/

# Code Availability

The code related to the key findings of this study will be available in the supplementary materials.

# Acknowledgments

This study was supported by National Science and Technology Major Project(Grant No. 2022ZD0116202)
We express gratitude to PubMed for providing accessible scientific literature data to the public.

# Author Contributions

Y.Z. conceived and designed the fundamental project structure, was involved in the initial development, and made significant contributions to system modifications.
D.W. collaborated with Y.Z. in constructing the first project version.
C.H. led the development of the current project version, authored the article, and is responsible for ongoing project maintenance.
Z.R. was responsible for the creation of specific web pages, enhancing webpage aesthetics, optimizing user interaction processes, and designing user guidance for website usage procedures.
Y.W. reviewed the paper and proposed many meaningful suggestions for revision
K.S. played a pivotal role in securing the necessary support that enabled the successful execution of this research representing the Center for Long-term Artificial Intelligence

# Competing Interests

The authors declare that there is no conflict of interest.